\documentclass[11pt]{article}
\usepackage[dvipdfmx]{graphicx}
\usepackage{%color,graphicx,amsmath,amssymb,
bm}

\newtheorem{prop}{Prop}[section]

\newtheorem{lemma}[prop]{Lemma}

\newtheorem{theorem}[prop]{Theorem}

\newtheorem{notes}[prop]{Note}

\usepackage{amsmath,amssymb}
\usepackage{makeidx}
\makeindex
\usepackage{ascmac}
\usepackage{pb-diagram}
\usepackage{here}
\usepackage{esint}
\title{Boundedness of Susceptibility in Spin Glass Transition 
of Transverse Field Mixed $p$-spin Glass Models 
}
\author{Chigak Itoi $^1$ and Yoshinori Sakamoto$^2$\\%}
\\
$^1$Department of Physics, GS $\&$ CST, Nihon University, \\
Kanda-Surugadai, Chiyoda, Tokyo 101-8308, Japan\\
\\
$^2$ Laboratory of Physics, CST, Nihon University, \\
Narashinodai, Funabashi-city, Chiba 274-8501, Japan}

\begin{document}
\maketitle
%
%\abst{
\begin{abstract}{
Spontaneous symmetry breaking phenomena in 
the transverse field mixed $p$-spin glass model in finite
dimensions are studied with Nishimori's  gauge theory. Useful identities in the gauge theory enable us to study 
$\mathbb Z_2$-symmetry breaking.
It is proven that ferromagnetic long-range order and spontaneous magnetization at an arbitrary temperature 
are bounded by those on the Nishimori line in the corresponding classical model.
These bounds imply that neither ferromagnetic long-range order nor spontaneous magnetization exists
in spin glass phase. It is proven also that
the ferromagnetic susceptibility has an upper bound in paramagnetic and spin glass phases.  
% \PACS{PACS code1 \and PACS code2 \and more}
% \subclass{MSC code1 \and MSC code2 \and more
}\end{abstract}
%}
%
\section{Introduction}    
Nishimori's gauge theory is a useful tool to understand properties of disordered spin systems \cite{N0,N}. 
The paramagnetic and ferromagnetic phases include the Nishimori line, where one can obtain 
several exact results of physical quantities, such as  the exact internal energy, bound on the specific heat  and identities among
correlation functions. These informations are quite helpful to understand the phase transition of
spin glasses in finite and infinite dimensions.
It is well-known that fluctuation of order parameters
is suppressed on the Nishimori line. The  absence of replica symmetry breaking
 on the Nishimori line has been argued in classical Ising spin glasses \cite{NS}.  

On the other hand, in quantum disordered spin systems, 
there are several important studies \cite{MON,I2,AGL,GAL,CGP,CL,Cr,IISS,W}.
They are less than those in classical Ising spin glasses, since 
the non-commutativity of operators yields considerable complication for evaluations.
Nonetheless, Nishimori's gauge theory is still useful for quantum  spin glasses to draw their phase diagrams \cite{MON}.
Bounds on some quantum correlation functions are evaluated in terms of the classical spin model on the Nishimori line. 
The absence of ferromagnetic long-range order on the Nishimori line in the classical Ising spin glasses implies 
that there is also no long-range order in the corresponding quantum systems. Although the Nishimori line lies out of the
spin glass phase, the gauge theory gives also helpful information of spin glass phase in quantum spin glass systems.

In the present paper, several identities on the Nishimori line are derived, and utilized them to study phase transitions in
quantum spin glass systems. 
These identities represent correlation functions in quantum systems at arbitrary temperatures 
in terms of those in the corresponding classical Ising systems on the Nishimori line. 
These representations enable us to prove well-known properties of the spontaneous $\mathbb Z_2$-symmetry breaking
in quantum spin glass systems. 
It is proven that there is neither ferromagnetic long-range order nor spontaneous magnetization in spin glass phase.
The square of magnetization and  spontaneous magnetization for any temperature 
 are bounded by those on the Nishimori line in the corresponding classical model without the transverse field.
 Several assumptions and Nishimori's gauge theory enable us to prove
the finiteness of ferromagnetic susceptibility in spin glass phase transition, which is observed also experimentally
 \cite{NKH}.
A bound on the ferromagnetic susceptibility is given by a correlation function 
in the corresponding classical model on the Nishimori line. 
Under acceptable assumptions, it is proven that 
the ferromagnetic susceptibility under zero external field does not diverge
in paramagnetic and spin glass phases at any lower temperature in quantum spin glass systems.

 %\section{Quantum  mixed $p$-spin glasses in finite dimensions}
 \section{Definitions}
For a positive integer $L$,  let  $\Lambda_L:= [0,L-1]^d \cap {\mathbb Z}^d
$ be a $d$ dimensional cubic lattice whose volume is $|\Lambda_L|=L^d$. 
Let  $Q \subset {\mathbb Z}$  be a finite set  of positive integers.
 For $p \in Q$, define  a collection ${\cal A}_p$ of
 interaction ranges $A_p \subset \Lambda_L$,  such that
$(0, \cdots, 0) \in  A_p$ and $|A_p| =p$.  
  Define a  collection ${\cal B}_p$ of interaction ranges  by
$$
{\cal B}_p:= \{X \subset \Lambda_L | X= i+ A_p, i\in \Lambda_L,  A_p \in {\cal A}_p  \}.
$$
 A sequence of spin operators  $(\sigma^{w}_i)_{w=x,y,z, i \in \Lambda_L}$
on a Hilbert space ${\cal H} :=\bigotimes_{i \in \Lambda_L} {\cal H}_i$ is
defined by a tensor product of the Pauli matrix $\sigma^w$ acting on ${\cal H}_i \simeq {\mathbb C}^{2}$ and unities.
These operators are self-adjoint and satisfy the commutation relations
$$
 [\sigma_k^y,\sigma_j^z]=2i \delta_{k,j} \sigma_j^x ,\ \ \  \ \ 
[\sigma_k^z,\sigma_j^x]=2i \delta_{k,j} \sigma_j^y ,\ \  \ \ \ [\sigma_k^x,\sigma_j^y]=2i \delta_{k,j} \sigma_j^z ,  
$$
and each spin operator satisfies
$$
(\sigma_j^w)^2 = {\bf 1}.
$$

Denote a product of  spins $$
\sigma_X^w= \prod_{i \in X} \sigma_i^w,
$$ 
for a finite sub-lattice $X \subset \Lambda_L$.
 Hamiltonian of mixed $p$-spin interactions with a transverse field $h>0$ is defined  by
\begin{equation}
H(\sigma, \bm J,h) :=-\sum_{p \in Q} \sum_{X \in {\cal B}_p}  J^p_X
\sigma_{ X}^z- h\sum_{i\in \Lambda_L}\sigma_i^x ,
\label{Hamilq}
\end{equation}
where,  a sequence 
$\bm J:=(J_X^p)_{X \in {\cal B}_p , p\in Q}$ consists  of  independent Gaussian 
random variables (r.v.s) with its expectation value  $\mu_p >0$ and its standard deviation $\Delta_p> 0$.
The probability density function of each $J_X^p$ is given by
\begin{equation}
P_p(J_X^p) := \frac{1}{\sqrt{2\pi \Delta_p^2}} \exp \Big[-\frac{(J_X^p-\mu_p)^2}{2 \Delta_p^2}\Big].
\label{PJ}
\end{equation}
${\mathbb E}$ denotes the sample expectation over 
all $J_X^p$, such that  
$$
{\mathbb E} J_X^p=\mu_p, \ \ \ {\mathbb E} (J_X^p -\mu_p)^2=\Delta_p^2.
$$
Gaussian r.v.s $J_X^p$ for 
$X \in {\cal B}_p , p\in Q$ are represented in terms of the independent and  identically distributed (i.i.d.) standard Gaussian
 r.v.s $g_X^p$
\begin{equation}
J_X^p = \Delta_p g_X^p+\mu_p.
\label{stanGauss}
\end{equation}
Note the $\mathbb Z_2$ symmetry
\begin{equation}
UH(\sigma, \bm J, h) U^\dag=H(\sigma, \bm J,h),
\end{equation}
for $U =\sigma_{\Lambda_L }^x$, 
if  all $p \in Q$ are even integers.

Define Gibbs state for the Hamiltonian.
For a positive $\beta $ and  real numbers $J_X^p$,  the  partition function is defined by
\begin{equation}
Z_L(\beta, \bm J, h) := {\rm Tr} e^{ - \beta H(\sigma,\bm J,h)},
\end{equation}
where the trace is taken over all basis in the Hilbert space. 
Let   $f$ be an arbitrary function 
of spin operators. The  expectation of $f$ in the Gibbs state is given by
\begin{equation}
\langle f(\sigma) \rangle^h_\beta
:=\frac{1}{Z_L(\beta, \bm J, h)}{\rm Tr} f( \sigma)  e^{ - \beta H(\sigma , \bm J, h )}.
\end{equation}
The index $h$ of the Gibbs expectation  represents the quantum perturbation in the Hamiltonian.
Note that the Gibbs expectation of operators $(\sigma_i^z)_{i\in \Lambda_L}$ is identical to 
at $h=0$ the Gibbs expectation  in the classical model
$$
\langle f(\sigma^z) \rangle^0_\beta =\langle f(\tau) \rangle_\beta,
$$ 
where $\tau:=(\tau_i)_{i\in \Lambda_L}$ is a sequence of eigenvalues of $(\sigma_i^z)_{i\in \Lambda_L}$.
Duhamel function of two functions $f(\sigma), g(\sigma)$ of spin operators
is defined by
\begin{equation}
(f(\sigma), g(\sigma))^h_\beta:=\int_0^1dt \langle e^{t\beta  H(\sigma, \bm J, h)}  f(\sigma)e^{-t \beta  H(\sigma, \bm J, h)} g(\sigma) \rangle^h_\beta
\end{equation}
Note that
$$
(f(\sigma^z), g(\sigma^z))^0_\beta =\langle f(\tau) g(\tau)\rangle_\beta,
$$ 
We define the following functions of  $(\beta,  \bm \Delta , \bm  \mu) \in [0,\infty)^{1+2|Q|} $ and randomness
$\bm J=(J_X^p)_{X \in {\cal B}_p, p \in  Q}$
\begin{equation}
\psi_L(\beta,\bm J,h) := \frac{1}{|\Lambda_L|} \log Z_L(\beta, \bm J,h), \\ 
\end{equation}
$-\frac{L^d}{\beta}\psi_L( \beta,\bm J,h)$ is called free energy in statistical physics.
Define a function $p_L:[0,\infty)^{1+2|Q|} \rightarrow {\mathbb R}$ by
\begin{eqnarray}
p_L(\beta, \bm \Delta, \bm \mu, h):={\mathbb E} \psi_L(\beta, \bm J,h ).
\end{eqnarray}

The following infinite volume limit
\begin{eqnarray}
p(\beta, \bm \Delta, \bm \mu, h):=
\lim_{L\to\infty}p_L(\beta, \bm \Delta, \bm \mu, h),
\end{eqnarray}
 exists for each $(\beta, \bm \Delta, \bm \mu, h)$ \cite{I2}.
Note that the function $\psi_L(\beta, \bm J,h)$, $p_L(\beta, \bm \Delta, \bm \mu, h)$  and $p(\beta, \bm \Delta, \bm \mu, h)$ 
are convex functions of each variable.

To study $\mathbb Z_2$-symmetry and its breaking, 
define an extended  $p$-th order operator in terms of $z$ component of the Pauli operators 
\begin{equation}
o_p := \frac{1}{|{\cal B}_p|} \sum_{X \in {\cal B}_p} \sigma_X^z.
\end{equation}
%{ \lemma \label{maglimt} The function $p(\beta, \Delta_1, \Delta_2, \mu_1,\mu_2 )$ 
% is continuously differentiable at almost all $\mu_1$, where the infinite volume limit exists and 
%$$
%\lim_{L\to\infty}\mathbb E   \langle o_1 \rangle_{\beta}^h
%= \frac{\partial }{\partial \mu_1} p(\beta, \Delta_1, \Delta_2, \mu_1, \mu_2, h),
%$$
%is valid,  since $p$ is a convex function of $\mu_1$ \cite{I2}. }

For arbitrary functions $f(\sigma), g(\sigma)$ of spin operators,
denote 
their truncated Duhamel correlation function by
$$
(f(\sigma);g(\sigma))_\beta^h:= (f(\sigma), g(\sigma))_\beta ^h- \langle f(\sigma)\rangle_\beta ^h\langle g(\sigma) \rangle_\beta ^h.$$
Note that the derivative of the expectation value is represented in terms of the truncated Duhamel function
$$
\frac{\partial}{\partial \mu_p} \langle f(\sigma) \rangle_\beta^h= \beta |{\cal B}_p| (f(\sigma);o_p)_\beta^h.
$$

\section{Nishimori's Gauge Theory}
Nishimori's gauge theory can be extended to mixed $p$-spin glass models \cite{IS}.
Define the Nishimori manifold (NM) by
\begin{equation}
\beta _p  \Delta_p^2 = \mu_p,
\end{equation}
for all $p\in Q$  in the  coupling constant space of $(\beta, \bm \Delta, \bm \mu)
\in [0, \infty) ^{1+2|Q|} $. 
Let us define a gauge transformation in Nishimori's gauge theory for spin glass \cite{CGN,MNC,N}.
For a spin configuration $\tau \in \{ 1,-1\} ^{ \Lambda_L}$, 
define a gauge transformation by
\begin{equation}
J_X^p \to  J_X^p \tau_X,   \ \ \  \sigma_X^z \to  \tau_X\sigma_X^z
=U(\tau)\sigma_X^zU(\tau)^{\dag},
\end{equation}

where $U(\tau):=\prod_{j\in\Lambda_L}(\sigma_j^x)^{(1-\tau_j)/2}$. 

The Hamiltonian is invariant under the gauge transformation.
\begin{equation}
H(\tau\sigma^z,\sigma^x, \bm J \tau, h ) = H(\sigma^z, \sigma^x, \bm J,h).
\label{gaugeinv}
\end{equation}
The distribution function is transformed in the following covariant form
\begin{equation}
P_p (J_X^p \tau_X)= P_p (J_X^p) e^{\frac{\mu_p}{\Delta_p^2} J_X^p (\tau_X-1) },
\label{gaugecov}
\end{equation}
%where $$ \tilde P_p (J_X^p) :=  \frac{1}{\sqrt{2\pi \Delta_p^2}} \exp \Big[-\frac{(J_X^p)^2+(\mu_p)^2}{2 \Delta_p^2} \Big].$$
Several of the following identities among correlation functions on the NM 
are shown in Ref.\cite{MON}. Some others are their extensions.

{\lemma \label{NMq} Denote $\beta _p := \mu_p/\Delta_p^2$ for all $p\in Q$ %, then $\beta$ is identical to $\beta_{\rm N}$
on the NM.
The one point function for $X \in {\cal B}_p$ satisfies
\begin{equation}
\mathbb E \langle \sigma_X^z \rangle_\beta^h=  \mathbb E \langle \sigma_X^z  \rangle_\beta^h \langle \tau_X  \rangle_{\beta _p  },
\label{1pointNMq}
\end{equation}
and two point  functions for $X,Y \in {\cal B}_p$  satisfy 
\begin{equation}
\mathbb  E \langle \sigma_X^z \rangle_\beta^h \langle  \sigma_Y^z \rangle_\beta^h =  \mathbb E \langle \sigma_X^z \rangle_\beta^h \langle 
\sigma_Y^z \rangle_\beta^h \langle \tau_X \tau_Y  \rangle_{\beta _p  },
\ \ \ \mathbb E \langle \sigma_X^z \sigma_Y^z \rangle_\beta^h =  \mathbb E
 \langle \sigma_X^z \sigma_Y^z   \rangle_\beta^h \langle \tau_X \tau_Y   \rangle_{\beta _p  }. \label{2pointNMq}
\end{equation}
Also Duhamel function and truncated Duhamel function
satisfy
\begin{equation}
 \mathbb E (\sigma_X^z, \sigma_Y ^z)_\beta^h =  \mathbb E
 ( \sigma_X^z, \sigma_Y^z   )_\beta^h  \langle \tau_X \tau_Y   \rangle_{\beta _p  }, \ \ \ \ 
 \mathbb E (\sigma_X^z; \sigma_Y ^z)_\beta^h =  \mathbb E
 ( \sigma_X^z; \sigma_Y^z   )_\beta^h  \langle \tau_X \tau_Y   \rangle_{\beta _p  }. \label{DuhamelNM}
\end{equation}
%Multiple point functions satisfy corresponding  extended formulae.\\
\\
Proof. } For 
$p \in Q$, $X \in {\cal B}_p$ and $\beta _p  := \frac{\mu_p}{\Delta_p^2}$, the 
 one point function $\mathbb E \langle \sigma_X^z\rangle_\beta^h$ written in the integration over
$\bm J$ can be represented in terms of gauge transformed form using the gauge invariance (\ref{gaugeinv}) of the Hamiltonian  
and the gauge covariance (\ref{gaugecov}) of the distribution 
\begin{eqnarray}
\mathbb E \langle \sigma_X^z \rangle _\beta^h &=& 
\int \langle \sigma_X^z  \rangle_\beta^h  \prod_{p\in Q}\prod_{X \in {\cal B}_p}P_p(J^p_X)dJ_X^p 
 =
  \int  \langle \sigma_X^z  \rangle_\beta^h \tau_X \prod_{p\in Q}\prod_{X \in {\cal B}_p} P_p(J^p_X\tau_X)  dJ_X^p
 %\mathbb E\nonumber \\
 %&=& \int \langle \sigma_X^z  \rangle_\beta^h \tau_X  P(J^p_X \tau_X) dJ_X^p \prod_{q\in Q, q\neq p}\prod_{X \in {\cal B}_q}P(J^q_X) dJ_X^q 
 %\nonumber \\
   \nonumber \\
  &=& \int \langle \sigma_X^z  \rangle_\beta^h \tau_X   \prod_{p\in Q}\prod_{X \in {\cal B}_p} P_p (J_X^p)  
  e^{\frac{\mu_p}{\Delta_p^2} 
   J_X^p (\tau_X-1)}dJ_X^p 
 \nonumber \\
  &=& 2^{-|\Lambda_L|}\sum_{\tau \in \{1,-1\}^{\Lambda_L}} \int \langle \sigma_X^z  \rangle_\beta^h \tau_X
   e^{\sum_{p\in Q}\sum_{X \in {\cal B}_p}\beta _p  J_X^p \tau_X }
\prod_{p\in Q}\prod_{X \in {\cal B}_q}  P_p (J_X^p)  e^{-\frac{\mu_p}{\Delta_p^2} 
   J_X^p }dJ_X^p  \nonumber \\
 &=& 2^{-|\Lambda_L|}\int \langle \sigma_X^z  \rangle_\beta^h  \langle  \tau_X  \rangle_{\beta _p  }
\sum_{\xi \in \{1,-1\}^{\Lambda_L}} e^{\sum_{p\in Q}\sum_{X \in {\cal B}_p}\beta _p  
J_X^p \xi_X } \prod_{p\in Q}\prod_{X \in {\cal B}_p}
P_p (J_X^p) e^{-\frac{\mu_p}{\Delta_p^2} 
   J_X^p } dJ_X^p \nonumber \\
  &=& 2^{-|\Lambda_L|}\sum_{\xi \in \{1,-1\}^{\Lambda_L}} \int \langle \sigma_X^z  \rangle_\beta^h  \langle  \tau_X  \rangle_{\beta _p  }
\prod_{p\in Q}\prod_{X \in {\cal B}_p} P_p (J_X^p\xi_X )  dJ_X^p
 \nonumber \\
  &=& 2^{-|\Lambda_L|}\sum_{\xi \in \{1,-1\}^{\Lambda_L}} \int \langle \sigma_X^z  \rangle_\beta^h  \langle  \tau_X  \rangle_{\beta _p  }
\prod_{p\in Q}\prod_{X \in {\cal B}_p} P_p (J_X^p)  dJ_X^p=\mathbb E \langle \sigma_X^z  \rangle_\beta^h  \langle  \tau_X  \rangle_{\beta _p  },
\label{1point-gauge}
\end{eqnarray}
where we have used the gauge
 invariance of % the one point function
$\langle \sigma_X^z  \rangle_\beta^h  
\langle  \tau_X  \rangle_{ \beta_p }$ under 
another gauge transformation 
$\sigma_X^z \to \xi_X\sigma_X^z, \  \tau_X \to \xi_X\tau_X ,\  J_X \to J^p_X\xi_X $ % has been used
 to obtain the last line.
Other identities of  multiple point functions are obtained in the same way. $\Box$  

{\notes It is possible that Nishimori's gauge theory is applicable to models with random interactions satisfying
other distributions, for example
the binomial distribution defined by
$$
P(J_X^p):= r_p \delta(J^p_X -\mu_p)+(1-r_p) \delta(J^p_X+\mu_p)
$$
for $0 \leq  r_p \leq 1$. In this distribution, the NM is defined by
$$
\beta_p  = K(r_p)
$$ 
for all $p\in Q$, where the function $K(r_p)$ is defined by
$$
%K_p^r 
 K(r_p) := \frac{1}{2} \log \frac{r_p}{1-r_p}.
$$
Throughout the text in the present paper, the Gaussian random $p$-spin interactions are treated for convenience and simplicity,  
since recent mathematical studies on spin glasses have been progressed mainly for Gaussian random interactions.
Several results have been obtained by combinations of Nishimori's gauge theory and recent useful studies \cite{IS}.  
Extensions of theorems in the present paper to those for the binomial distribution
are straightforward.}

\section{$\mathbb Z_2$-symmetry Breaking in the Spin Glass Transition}
Here, we study spin glass phase transition in the transverse field mixed even $p$-spin glass model.
 Let $Q_2$ be a finite set  of positive even integers, and define $Q:=\{1\} \cup Q_2$  
and consider the transverse field mixed even $p$-spin glass model with $Q$ 
 and   coupling constants $(\beta, \bm \Delta, \bm \mu, h)$.  
 Note that the perturbation $(\Delta_1, \mu_1)\neq (0,0)$ 
breaks the $\mathbb Z_2$-symmetry. The operator $o_1$ is the ferromagnetic order operator, which measures the $\mathbb Z_2$-symmetry breaking.

\subsection{Absence of long-range order and spontaneous magnetization}
First, absence of spontaneous ferromagnetic  magnetization is proven in addition to that of 
ferromagnetic long-range order, already indicated by Morita, Ozeki and Nishimori \cite{MON}.

{\theorem \label{corq}  
Consider  the transverse field mixed even $p$-spin glass 
model with $Q=\{1\}\cup Q_2$  for an arbitrary $\beta >0$, and 
define  NM by $\mu_p/\Delta_p^2$ taking the same value for all $p\in Q_2$.
If there is no ferromagnetic long-range order for 
$(\beta=\mu_p/\Delta_p^2, \Delta_1=0,\Delta_p, \mu_1=0, \mu_p,h=0)$ on the NM in the corresponding classical spin glass model,
then there is no ferromagnetic long-range order \begin{equation}
\lim_{L\to \infty} \mathbb E   \langle o_1^2 \rangle_{\beta}^h =0.
\end{equation}
 for any
$(\beta, \Delta_1=0,  \Delta_p,
\mu_1=0, \mu_p, h\neq0)$  in the model with the quantum perturbation by the transverse field $h$.

If there is no spontaneous ferromagnetic magnetization
for  $(\beta=\mu_p/\Delta_p^2, \Delta_1, \Delta_p, \mu_1, \mu_p,h=0)$ on the NM  in 
the corresponding classical spin glass model,  
then there is no spontaneous ferromagnetic magnetization 
\begin{equation}
\lim_{\mu_1\to 0}\lim_{L\to \infty} \mathbb E   \langle o_1 \rangle_{\beta}^h
 =0.
\end{equation}
 for  any
$(\beta, \Delta_1, \Delta_p, \mu_1, \mu_p, h\neq0)$ in the model with 
the quantum perturbation by the transverse field $h$.
\\

\noindent
Proof.}  These are proven using Lemma \ref{NMq}. First, we prove absence of long-range order.
 The identity (\ref{2pointNMq}) and  Jensen's inequality imply
 \begin{eqnarray}
\mathbb E  \langle o_1^2 \rangle_{\beta} ^h
&=&\frac{1}{|\Lambda_L|^2} \sum_{i,j\in \Lambda_L}  \mathbb  E \langle \sigma_i^z \sigma_j^z  \rangle_{\beta}^h
= \frac{1}{|\Lambda_L|^2} \sum_{i,j\in \Lambda_L} \mathbb  E \langle \sigma_i^z \sigma_j^z  \rangle_{\beta}^h
   \langle \tau_i \tau_j  \rangle_{\beta_{\rm N}} \nonumber \\
  &\leq&   \frac{1}{|\Lambda_L|^2} \sum_{i,j\in \Lambda_L} \mathbb  E |\langle \sigma_i^z \sigma_j^z  \rangle_{\beta}^h||
   \langle \tau_i \tau_j  \rangle_{\beta_{\rm N}} | \leq   
   \frac{1}{|\Lambda_L|^2} \sum_{i,j\in \Lambda_L} \mathbb  E   |\langle \tau_i \tau_j  \rangle_{\beta_{\rm N}} |\nonumber \\
&\leq&  \frac{1}{|\Lambda_L|^2} \sum_{i,j\in \Lambda_L}\sqrt{ \mathbb  E   \langle \tau_i \tau_j  \rangle_{\beta_{\rm N}}^2}
\leq
 \sqrt{ \frac{1}{|\Lambda_L|^2} \sum_{i,j\in \Lambda_L} \mathbb  E   \langle \tau_i \tau_j  \rangle_{\beta_{\rm N}}^2}\nonumber \\
&=& \sqrt{ \frac{1}{|\Lambda_L|^2} \sum_{i,j\in \Lambda_L} \mathbb  E   \langle \tau_i \tau_j  \rangle_{\beta_{\rm N}}}=
\sqrt{ \mathbb  E   \langle \Big( \frac{1}{|\Lambda_L|} \sum_{i\in \Lambda_L} \tau_i  \Big)^2 \rangle_{\beta_{\rm N}} },
\end{eqnarray}
where the identity (\ref{2pointNMq}) for $\beta_p=\beta_{\rm N}$ for all $p\in Q_2$ 
has been used to obtain the last line. 
Therefore,  the assumption implies no long-range order also for  for any
$(\beta, \Delta_1=0,  \Delta_p,
\mu_1=0, \mu_p, h\neq0)$ 
\begin{equation}
\lim_{L\to \infty} \mathbb E  \langle o_1^2 \rangle_{\beta}^h =0.
\end{equation}
Next, let us consider a magnetization process for 
$(\beta,\Delta_1, \Delta_p, \mu_1, \mu_p, h\neq 0) $. If we denote $\beta_{\rm N}:= (\mu_1/\Delta_1^2, \mu_p/\Delta_p)$, 
 the identity (\ref{1pointNMq}) gives a bound on
 the  magnetization. 
\begin{equation}
|{\mathbb E} \langle \sigma_i^z \rangle_{\beta}^h|
=|{\mathbb E} \langle \sigma_i^z \rangle_{\beta}^h\langle \tau_i \rangle_{\beta_{\rm N} } |
\leq {\mathbb E}| \langle \sigma_i^z \rangle_{\beta}^h ||\langle \tau_i \rangle_{\beta_{\rm N} } |
\leq {\mathbb E}|\langle \tau_i \rangle_{\beta_{\rm N} } |
\leq \sqrt{{\mathbb E} \langle \tau_i \rangle_{\beta_{\rm N} }^2}
 =\sqrt{{\mathbb E} \langle \tau_i \rangle_{\beta_{\rm N} }},
\end{equation}
for any $i \in \Lambda_L$. This and  Jensen's inequality imply
\begin{equation}
\mathbb E   \langle o_1 \rangle_{\beta}^h
=\frac{1}{|\Lambda_L|} \sum_{i\in \Lambda_L} 
{\mathbb E} \langle \sigma_i ^z\rangle_{\beta}^h \leq 
 \frac{1}{|\Lambda_L|} \sum_{i\in \Lambda_L} 
\sqrt{{\mathbb E} \langle \tau_i \rangle_{\beta_{\rm N}}} \leq 
\sqrt{ \frac{1}{|\Lambda_L|} \sum_{i\in \Lambda_L} 
{\mathbb E} \langle \tau_i \rangle_{\beta_{\rm N}}}.
\end{equation}
%Note that the limit $\mu_1 \to 0$ implies $\beta_1 \to 0.$ 
Therefore, the assumption on the spontaneous magnetization
on the NM implies that
there is no spontaneous magnetization for an arbitrary $\beta >0$
\begin{equation}
\lim_{\mu_1\to 0}\lim_{L\to \infty} \mathbb E   \langle o_1 \rangle_{\beta}^h
 =0.
\end{equation}
This completes the proof. $\Box$

\subsection{Bound on the susceptibility}
Finally, let's prove that the ferromagnetic susceptibility has a finite upper bound  under 
acceptable assumptions in the $\mathbb Z_2$-symmetry breaking 
phase transition of  the transverse field  mixed even $p$-spin glass model.  
For $\Delta_1=0$, 
regard the sample expectation of the magnetization as a function of the deterministic longitudinal field $\mu_1$ and the system size $L$.
Define a function
\begin{equation}
m_L(\mu_1) :=\frac{1}{\beta} \frac{\partial}{\partial  \mu_1} p_L(\beta, \Delta_1=0, \Delta_p, \mu_1,\mu_p , h)
=\mathbb E \langle o_1 \rangle_\beta^h
=\frac{1}{|\Lambda_L|} \sum_{i\in \Lambda_L} \mathbb E\langle  \sigma_i^z \rangle_\beta ^h.
\end{equation}

\paragraph{Assumptions }{\it  Consider the  transverse field mixed even $p$-spin glass
model with $Q=\{1\} \cup Q_2$ for arbitrary coupling constants $(\beta, \Delta_1=0 , \Delta_p,\mu_1, \mu_p, h)$. 
Assume the following  A1 and A2.\\
\\
A1. 
There exists a positive number $C$ independent of $L$, such that
\begin{equation} 
 \frac{1}{|\Lambda_L|} \sum_{i,j\in \Lambda_L} \sqrt{ \mathbb  E \langle \tau_i \tau_j  \rangle_{\beta_{\rm N}}}
\leq C,
\end{equation}
 for   $(\beta,\Delta_1= 0 , \Delta_p,\mu_1, \mu_p, h=0)$ on the  NM defined by $\beta_{\rm N} := \mu_p/\Delta_p^2$.  
   % in the paramagnetic phase of  the corresponding classical spin glass model.
 \\
\\
A2.  The finite size ferromagnetic susceptibility 

which is defined by $m_L'(\mu_1):=\frac{\partial}{\partial\mu_1}m_L(\mu_1)$

 is bounded by 
\begin{equation}
m_L'(\mu_1) \leq m_L'(0)
\end{equation}
for sufficiently small $|\mu_1| \neq 0$ for sufficiently large $L$. 
}\\
%Here, we discuss acceptability of A1 and A2. 
\\
A1 is valid, if the function $\mathbb E \langle \tau_i \tau_j \rangle_{\beta_{\rm N}}$  decays exponentially
for $|i-j| \gg 1$ %This is guaranteed by the exponential decay of the function $\mathbb E \langle \tau_i \tau_j \rangle_{\beta_{\rm N}}$ 
on the NM in the paramagnetic phase of the classical  mixed $p$-spin glass  model for $h=0, \Delta_1=0, \mu_1=0$. 
%Since  the identity (\ref{2pointNMq}) and A1 imply 
%$$
%\frac{1}{|\Lambda_L|^2} \sum_{i,j\in \Lambda_L} \mathbb  E \langle \tau_i \tau_j  \rangle_{\beta_{\rm N}}
%=\frac{1}{|\Lambda_L|^2} \sum_{i,j\in \Lambda_L}  \mathbb  E \langle \tau_i \tau_j  \rangle_{\beta_{\rm N}}^2
%\leq 
%\frac{1}{|\Lambda_L|^2} \sum_{i,j\in \Lambda_L} \sqrt{ \mathbb  E \langle \tau_i \tau_j  \rangle_{\beta_{\rm N}}}
%\leq \frac{C}{|\Lambda_L|},
%$$ 
Note that A1 provides a sufficient condition for 
the absence of long-range order in the paramagnetic phase of the classical model assumed in Theorem \ref{corq}.
\\
A2 implies that the singular behavior of the susceptibility for $\mu_1\neq0$ becomes weaker  than that for $\mu_1=0$.
%is valid, if $m_L'''(0) \leq 0$ which is equivalent to $\mathbb E [( o_L,o_L,o_L,o_L)_\beta^h -3{(o_L,o_L)_\beta^h}^2] \leq 0$ for $\mu_1=0$.
%This implies $m_L'(0) \geq m_L'(\mu_1)$ for sufficiently weak $\mu_1$,since the $\mathbb Z_2$-symmetry gives $m_L''(0) = 0$. 

{\theorem \label{sus}  
Consider the transverse field  mixed even $p$-spin glass
model with $Q=\{1\} \cup Q_2$ for arbitrary coupling constants $(\beta, \Delta_1=0 , \Delta_p,\mu_1, \mu_p, h)$ 
, where above assumptions A1 and A2 are satisfied.
 Then, the ferromagnetic  susceptibility at $\mu_1=0$
 is bounded from the above 
 \begin{equation}
 \limsup_{\mu_1 \to 0}  \frac{\partial }{\partial \mu_1}\lim_{L\to\infty}  m_L(\mu_1)
 %\frac{1}{|\Lambda_L|} \sum_{i\in \Lambda_L} \mathbb E\langle  \sigma_i^z \rangle_\beta ^h
 %=\lim_{\mu_1 \to 0}  \frac{\partial }{\partial \mu_1}\limsup_{L\to\infty} 
 % \frac{\partial }{\partial \mu_1} p_L(\beta, 0, \Delta_2, \mu_1, \mu_2, h)
\leq 2\beta C.
\end{equation} 
also for   $(\beta, \Delta_1=0 , \Delta_p,\mu_1, \mu_p, h)$. 
in the infinite-volume limit.
\\

\noindent
Proof.} 
The function $p(\beta, \Delta_1,  \Delta_p, \mu_1,\mu_p  )$ 
 is continuously differentiable at almost all $\mu_1$, where the infinite volume limit of the ferromagnetic magnetization 
 exists and it is represented in terms of the partial derivative of $p$ in $\mu_1$
\begin{equation}
%m(\mu_1):=
\lim_{L\to\infty}m_L(\mu_1)
= \frac{1}{\beta}\frac{\partial }{\partial \mu_1} p(\beta, \Delta_1=0, \Delta_p, \mu_1, \mu_p, h),
\end{equation}
since $p(\beta, \Delta_1=0, \Delta_p, \mu_1, \mu_p, h)$ exists as a convex function of $\mu_1$ \cite{I2}.
The mean value theorem implies that for an arbitrary real number $\mu_1\neq 0 $, there exists a positive number $ \theta < 1$,  such that  
\begin{equation}
\frac{m_L(\mu_1)}{\mu_1}=
\frac{1}{\mu_1}[m_L(\mu_1) -m_L(0) ]= m_L'(\theta \mu_1).
\end{equation}
Note that the $\mathbb Z_2$-symmetry guarantees $m_L(0)=0$ for any $L$. 
The assumption A2 implies that the right hand side in the above identity is bounded by
\begin{equation}
m_L'(\theta \mu_1) \leq m_L'(0),
\end{equation}
 for a sufficiently weak
deterministic longitudinal field $\mu_1 \neq0$ and for sufficiently large $L$.
For $\mu_1=0$, 
the following bound on the susceptibility is obtained  
using the  identity
 (\ref{DuhamelNM}), Jensen's inequality and identity (\ref{2pointNMq}) for $\beta=\beta_{\rm N}$ 
\begin{eqnarray}
&&
m_L'(0)
=\frac{\beta}{|\Lambda_L|}\sum_{i,j\in \Lambda_L}\mathbb  E %[
( \sigma_i^z; \sigma_j^z )_\beta^{h}
 =\frac{\beta}{|\Lambda_L|}\sum_{i, j\in \Lambda_L}  \mathbb  E 
 ( \sigma_i ^z; \sigma_j^z )_\beta^{h} 
 \langle \tau_i \tau_j  \rangle_{\beta_{\rm N}} 
 \nonumber \\
&&\leq \frac{\beta}{|\Lambda_L|} \sum_{i,j\in \Lambda_L}  \mathbb  E| ( \sigma_i^z; \sigma_j ^z)_\beta^{h}
 || \langle \tau_i \tau_j  \rangle_{\beta_{\rm N}}| 
 \leq \frac{2\beta}{|\Lambda_L|} \sum_{i,j\in \Lambda_L}  \mathbb  E| \langle \tau_i \tau_j  \rangle_{\beta_{\rm N}}| \nonumber \\
 &&\leq\frac{2\beta}{|\Lambda_L|} \sum_{i,j\in \Lambda_L}\sqrt{  \mathbb  E \langle \tau_i \tau_j  \rangle_{\beta_{\rm N}}^2}
 %=
  = \frac{2\beta}{|\Lambda_L|} \sum_{i,j\in \Lambda_L}\sqrt{  \mathbb  E \langle \tau_i \tau_j  \rangle_{\beta_{\rm N}}}.
 \label{susq}
\end{eqnarray}
The assumption A1 on the correlation function at $ (\beta_{\rm N}, \Delta_1=0, \Delta_p, \mu_1=0, \mu_p, h=0)$ implies 
\begin{eqnarray}
m_L'(\theta \mu_1) \leq m_L'(0)
  \leq \frac{2\beta}{|\Lambda_L|} \sum_{i,j\in \Lambda_L}\sqrt{  \mathbb  E \langle \tau_i \tau_j  \rangle_{\beta_{\rm N}}}
  \leq 2 \beta C,
\end{eqnarray}
where $C$ does not depend on $L$. 
Therefore, 
\begin{eqnarray}
\lim_{L\to\infty}\frac{1}{\mu_1}[m_L(\mu_1) -m_L(0) ] \leq
\limsup_{L\to\infty}
 \frac{2\beta}{|\Lambda_L|} \sum_{i,j\in \Lambda_L}\sqrt{  \mathbb  E \langle \tau_i \tau_j  \rangle_{\beta_{\rm N}}}
  \leq 2 \beta C,
\end{eqnarray}
for sufficiently small $|\mu_1|$.
The susceptibility at any  $ (\beta, \Delta_1=0, \Delta_p, \mu_1=0, \mu_p, h)$
 is bounded from the above
\begin{equation}
\limsup_{\mu_1 \to 0}  \frac{\partial }{\partial \mu_1}\lim_{L\to\infty}m_L(\mu_1)
 = \limsup_{\mu_1 \to 0} \lim_{L\to\infty} \frac{1}{\mu_1}[m_L(\mu_1) -m_L(0) ] 
 \leq 2\beta C.
\end{equation}    
This completes the proof.   
 $\Box$
 
 {\notes 
 The results in Theorem \ref{corq} and Theorem \ref{sus}
  are well-known general properties of spin glasses, which are valid also in the classical systems given by $h=0$. 
 These results are  consistent with rounding effects obtained in Ref. \cite{AGL, AW, GAL}. The finiteness of susceptibility is 
observed experimentally in spin glasses  \cite{NKH}.
Nishimori's  gauge theory is useful also for  understanding experimental results in disordered spin systems. 
}\\
\\
\noindent
{\bf Acknowledgments} \\
 C.I. is supported by JSPS (21K03393).

\end{document}